\documentclass[12pt]{article}
\usepackage{graphicx}
\usepackage{amsmath}
\usepackage{amssymb}
\usepackage{caption2}
\begin{document}
\bibliographystyle{prsty}
\begin{center}
{\large {\bf \sc{  Reanalysis the pentaquark $\Theta^+(1540)$ in the framework of QCD sum rules approach with direct instantons }}} \\[2mm]
Zhi-Gang Wang$^{1}$ \footnote{Corresponding author; E-mail,wangzgyiti@yahoo.com.cn.  }, Wei-Min Yang$^{2}$ and Shao-Long Wan$^{2} $    \\
$^{1}$ Department of Physics, North China Electric Power University, Baoding 071003, P. R. China \\
$^{2}$ Department of Modern Physics, University of Science and Technology of China, Hefei 230026, P. R. China \\
\end{center}

\begin{abstract}
In this article, we study the pentaquark state $ \Theta^+(1540)$
with a (scalar)diquark-(pseudoscalar)diquark-antiquark type
interpolating current in the framework of the QCD sum rules approach
by including the contributions from the direct instantons. The
numerical results indicate that the contributions from the direct
instantons are very small  and can be safely neglected.
\end{abstract}


\section{Introduction}

In 2003, several collaborations have reported the observation of a
new baryon state $\Theta^+(1540)$ with positive strangeness and
minimal quark contents $udud\bar{s}$ \cite{exp2003}. The existence
of such an exotic state with narrow width $\Gamma < 15 MeV$ and
$J^P={\frac{1}{2}}^+$ was first predicted by Diakonov, Petrov and
Polyakov  in the chiral quark soliton model, where the
$\Theta^+(1540)$ is a member of the baryon antidecuplet
$\overline{10}$ \cite{Diakonov97}. The discovery  has opened a new
field of strong interaction  and provides a new opportunity for a
deeper understanding of the low energy QCD. Intense theoretical
investigations have been motivated
 to clarify the quantum numbers and to understand the
under-structures of the pentaquark state $\Theta^+(1540)$
\cite{ReviewPenta}. The zero of the third component of isospin
$I_3=0$ and the absence of isospin partners suggest  that the baryon
$\Theta^+(1540)$ is an isosinglet, while the spin and parity have
not been experimentally determined yet and  no consensus has ever
been reached  on the theoretical side. The extremely narrow width
below $10MeV$ puts forward a serious challenge to all theoretical
models, in the conventional  uncorrelated quark models the expected
width is of the order of several hundred $MeV$, since the strong
decay $\Theta^{+} \, \rightarrow K^+ N$ is Okubo-Zweig-Iizuka (OZI)
super-allowed.

Instantons, as the solutions of the classical Yang-Mills equation of
motion, play a crucial role in  description of the low energy strong
interactions, such as the $U(1)_{A}$ problem, dynamical chiral
symmetry breaking, tunneling the $\theta$ vacuum and so on
\cite{Schafer96}.
  In the quark-quark
sector, the instantons induced  't Hooft interaction has strong
flavor and spin dependence,  which can  explain a lot of hadronic
phenomena. The instanton induced effective lagrangian
 leads to a strong attractive interaction  in the color antitriplet
 channel $\overline{3}$ with $J^{P}=0^{+}$ which favors the
formation of  scalar  diquarks (such correlation may also arise from
the color-spin force of the one-gluon exchange), and a strong
repulsive interaction in the $0^-$ channel
\cite{Shifman80,GluonInstanton}. The instanton induced interactions
dominate the dynamics between quarks at intermediate distances about
$ \rho_c\approx \frac{1}{3} ~{\rm fm}$, which is much smaller than
the confinement size $R \approx 1~{\rm fm}$, therefore the quarks
may cluster together to form diquark or triquark in the confining
region. So it is interesting to investigate the contributions from
the direct instantons \cite{Lee04} \footnote{In writing the article,
the Ref.\cite{Lee04} appears, it is an interesting article.}.

In this article, we take the point of view that the quantum numbers
of the pentaquark $\Theta^{+}(1540)$ are $J=\frac{1}{2}$ , $I=0$ ,
$S=+1$, and study its mass with a
(scalar)diquark-(pseudoscalar)diquark-antiquark type interpolating
current in the framework of the QCD sum rules approach by including
the contributions from the direct instantons
\cite{sugiyama04,Shifman79}.

 The article is arranged as follows:  we introduce the instanton liquid model in
section II; in section III, we derive the QCD sum rules for the
pentaquark  state $\Theta^+(1540)$ with the contributions from the direct
instantons; in section IV, numerical results; section V is
reserved for conclusion.

\section{Instanton Liquid Model}
The instanton liquid model  is based on a semiclassical
approximation, in which  all gauge configurations are replaced by an
ensemble of topologically non-trivial fields i.e. instantons and
anti-instantons \cite{Schafer96,Ilgenfritz81}.  To avoid the
notorious infrared problem due to the large size instantons,  we can
suppose that  for larger distance, the vacuum gets more filled with
the instantons of increasing size, at some scale there might be some
repulsive interactions
  to stabilize the ensemble while  the semiclassical treatment is still
  possible and the instantons are not much deformed through the
  interactions,  thus form a dilute instanton liquid. Although it does
not give rise to a long range confining force between quarks, the
instanton vacuum has been shown to provide a good phenomenological
description of many hadronic properties, for example, the coefficients of the Chiral
 lagrangian \cite{Wang02}.
 Phenomenological, numerical  and lattice calculations show that
their total density is about $\bar{n}\simeq 1 {\rm fm}^{-4}$ while
the typical size is about $\rho_c\sim \frac{1}{3} {\rm fm}$,
leading to a small diluteness parameter $\bar{n}\rho^3\sim
10^{-2}$
 .   As the instanton vacuum
is fairly dilute,  we can take the single instanton approximation
for the collective effects in mathematical manipulation, which has
an outstanding advantage that we can carry out the calculations
analytically. In the single instanton approximation, the collective
contributions of all
 instantons other than the leading one
are taken into account  by a single effective parameter, the
effective mass $m^*$,
\begin{equation}
 m^*_q= m_q - \frac{2}{3}\pi^2\,\rho^2 \,<\bar{q}\,q>,
 \end{equation}
 which leads to
the value  $m^*\simeq 170 MeV$ for the u and d quarks while a
detailed  updated analysis suggests the  value $m^*\simeq 86 MeV$
\cite{Shuryak83}. In this article, we take the usually used value $m^*\simeq 170 MeV$.

The crucial property of instantons, originally discovered by 't
Hooft , is  the zero mode of the Dirac operator $iD\!\!\!\!/$   in
the instanton background,
\begin{eqnarray}
i D\!\!\!/\psi_0(x)& =& 0, \nonumber\\
\psi_{0\,\,a\,\nu}(x;z)& =& \frac{\rho}{\pi}
\frac{1}{\sqrt{(x-z)^2+\rho^2}^{3}}\left[ \frac{1 -
\gamma_5}{2}\, \frac{x\!\!\!/-z\!\!\!/}{\sqrt{(x-z)^2}}
\right]_{\alpha\,\beta} U_{a\,b}\,\epsilon_{\beta\,b},
\end{eqnarray}
 where $z$ denotes the instanton position, $\alpha,
\beta= 1,\cdots 4$ are spinor indices and $U_{a b}$ represents
color orientations.

Isolating the contributions from the zero-modes,  the quark
propagator in the instanton background  can be  written  as,
\begin{eqnarray}
 S_I(x,y;z) & =
&\frac{\psi_0(x-z)\,\psi^\dagger_0(y-z)}{i\,m} +
    \sum_{\lambda\ne 0} \, \frac{\psi_\lambda (x-z)\,
     \psi^\dagger_\lambda(y-z)}{\lambda + i\, m}\nonumber\\
    & = & S^{zm}_I(x,y;z) + S^{nzm}_I(x,y;z)  \; ; \nonumber\\
 S^{zm}_I(x,y;z)&=&
\frac{(x\!\!\!/-z\!\!\!/ )\gamma_\mu \gamma_\nu
(y\!\!\!/-z\!\!\!/)}{8 m} \left[ \tau_\mu^- \tau_\nu^+
\,\frac{1-\gamma_5}{2} \right] \, \phi(x-z)\, \phi(y-z),
\end{eqnarray}
where
 \begin{eqnarray}
\phi(t)=\frac{\rho}{\pi}\frac{1}{\sqrt{t^2}\,\sqrt{t^2+\rho^2}^{3}
}, \qquad \tau_\mu^{\pm} = ({\bf \tau}, \mp i   ) .\nonumber
 \end{eqnarray}
 In the chiral limit, $m\rightarrow 0$, the
$S^{nzm}_I(x,y;z)$  is   known exactly \cite{Brown78}. In the
  small distances limit $|x-y| \rightarrow 0$, or in extreme dilute limit $|x-z|\rightarrow \infty$,
  we can approximate the nonzero models  by $
  S^{nzm}_I(x,y;z) \simeq S_0(x,y)$,  with $S_0$ denotes the free propagator.

 In this article, the
 instanton liquid model is taken into account by the zero-mode part
 of  the single instanton approximation mathematically, i.e. $m\rightarrow m^*$ and  $S_I(x,y;z)\approx S^{zm}_I(x,y;z)$ .

 The corresponding  quark propagator for the anti-instanton can be
obtained through the substitution,
 \begin{eqnarray}
\frac{1-\gamma_5}{2}\longleftrightarrow \frac{1+\gamma_5}{2},
\, \,  \tau^- \longleftrightarrow \tau^+.
\end{eqnarray}

\section{QCD Sum Rules for the Pentaquark state $\Theta^+(1540)$ with Direct Instantons }
In the following, we study the  mass of the pentaquark state
$\Theta^+(1540)$ with the QCD sum rules approach  by including the
contributions from the direct instantons. Firstly, let us write down
the two-point correlation function,
\begin{equation}
 \Pi (p) =  i\int d^4 x\,  e^{ip\cdot x}  \langle 0| T[J(x) \bar J(0)
 ]|0\rangle,
  \end{equation}
  with
\begin{eqnarray}
  J(x)&=&\epsilon^{abc}\epsilon^{def}\epsilon^{cfg}
   \{u_a^T(x)Cd_b(x)\}\{u_d^T(x)C\gamma_5 d_e(x)\}C\bar{s}_g^T(x) , \nonumber \\
  \bar{J}(x)&=&-\epsilon^{abc}\epsilon^{def}\epsilon^{cfg}
   s_g^T(x)C\{\bar{d}_e(x)\gamma_5 C \bar{u}_d^T(x)\}\{\bar{d}_b(x)C \bar{u}_a^T(x)\} ,\nonumber
  \end{eqnarray}
here $a,b,c,\cdots$ are color indices and $C=i\gamma^2\gamma^0$
\cite{sugiyama04}.

According to the basic assumption of current-hadron duality in the
QCD sum rules  approach \cite{Shifman79}, we insert  a complete
series of intermediate states satisfying the unitarity   principle
with the same quantum numbers as the current operator $J(x)$
 into the correlation function in
Eq.(5)  to obtain the hadronic representation. After isolating the
pole term of the lowest pentaquark  state, we obtain the  result,
 \begin{eqnarray}
 \Pi(p) &= &  \lambda^2 \frac{\gamma \cdot p +m_{\Theta^+}}{m^2_{\Theta^+}-p^2}
 + \cdots \, ,
 \end{eqnarray}
here the  following definition has been used,
 \begin{equation}
 \langle 0| J(0) |B(p) \rangle= \lambda u( p) .
  \end{equation}
In the following, we perform the operator product expansion
 to obtain the spectral representation at the level of
quark and gluon degrees of freedom with the contributions from the
direct instantons.  As the instantons are solutions of the classical
Yang-Mills  equations in the Euclidean space-time,  we have to
rotate all the variables from the Minkowski  space-time region to
the Euclidian space-time region,
\begin{eqnarray}
 \Pi (p) &=& - \epsilon^{abc}\epsilon^{def}\epsilon^{cfg}\epsilon^{a'b'c'}\epsilon^{d'e'f'}\epsilon^{c'f'g'}\int_E d^4 x\,  e^{-ip\cdot x} \nonumber\\
 &&
 Tr\left\{CS_{bb'}(x)CS_{aa'}^T(x)\right
 \}Tr\left\{C\gamma_5S_{ee'}(x)C\gamma_5S_{dd'}^T(x)\right\}C{S_s}_{gg'}^T(-x)C,
  \end{eqnarray}
here the subscript $s$ denotes  the s quark.  The quark propagator
has two terms, the standard one ($st$) and the one in the
instanton background ($in$),
\begin{equation}
S_{ab}(x,y)=S^{st}_{ab}(x,y)+S^{in}_{ab}(x,y),
\end{equation}

In this article, we take into account the contributions from the
direct instantons by the zero modes in the single instanton
approximation for the instanton liquid model,
\begin{equation}
S^{in}_{ab}(x,y)\approx S_I^{zm} ,  \, m\rightarrow m^* \, .
\end{equation}
Substitute the above quark propagator in Eq.(9) for those in Eq.(8),
we can obtain the following result,
\begin{eqnarray}
 \Pi (p)&=& - \epsilon^{abc}\epsilon^{def}\epsilon^{cfg}\epsilon^{a'b'c'}\epsilon^{d'e'f'}\epsilon^{c'f'g'}\int_E d^4 x\,  e^{-ip\cdot x} \nonumber\\
 && \left\{
 Tr\left\{CS^{st}_{bb'}(x)C{S^{st}_{aa'}}^T(x)\right
 \}Tr\left\{C\gamma_5S^{st}_{ee'}(x)C\gamma_5{S^{st}_{dd'}}^T(x)\right\}
  C{{S_s}^{st}_{gg'}}^T(-x)C \right. \nonumber\\
 &+&
 Tr\left\{CS^{in}_{bb'}(x)C{S^{in}_{aa'}}^T(x)\right
 \}Tr\left\{C\gamma_5S^{st}_{ee'}(x)C\gamma_5{S^{st}_{dd'}}^T(x)\right\}C{{S_s}^{st}_{gg'}}^T(-x)C \nonumber\\
 &+&
  Tr\left\{CS^{st}_{bb'}(x)C{S^{st}_{aa'}}^T(x)\right
 \}Tr\left\{C\gamma_5S^{in}_{ee'}(x)C\gamma_5{S^{in}_{dd'}}^T(x)\right\}C{{S_s}^{st}_{gg'}}^T(-x) C \nonumber\\
 &+&\left.
 Tr\left\{CS^{in}_{bb'}(x)C{S^{in}_{aa'}}^T(x)\right
 \}Tr\left\{C\gamma_5S^{in}_{ee'}(x)C\gamma_5{S^{in}_{dd'}}^T(x)\right\}C{{S_s}^{st}_{gg'}}^T(-x)C \right\} . \nonumber\\
  \end{eqnarray}
  The important selection rule for the quarks in the instanton
  background
\begin{equation}
\stackrel{\rightarrow}{\sigma_i}\bigoplus \vec \tau_i=0,
\end{equation}
with $\sigma_i$ is usual spin and $\tau_i$ is color spin,  leads to
the vanishing of one-body (i.e. $S^{in}$), three-body, five-body
instanton induced contributions, and remaining  only the terms in
Eq.(11)\footnote{In this article, we study the contributions from
the direct instantons with the instanton liquid model,  as the
instanton vacuum is fairly dilute, we can take into account  the
collective effects of the instanton ensemble with the single
instanton approximation mathematically. If there is  just one
instanton,
 the last term in Eq.(11) should
vanish due to the Fermi statistics, however, we are dealing with the
dilute instanton liquid, the induced contributions  of all the $u$,
$d$, $s$ quarks in the dilute instanton ensemble should be taken
into account, in practical manipulation, we can choose the
corresponding ones with the single instanton approximation
mathematically. The diquarks $S^a(x) = \epsilon^{abc}
u_b^T(x)C\gamma_5 d_c(x),\epsilon^{abc} u_b^T(x)C\gamma_5
s_c(x),\epsilon^{abc} d_b^T(x)C\gamma_5 s_c(x) $ and $P^a(x) =
\epsilon^{abc} u_b^T(x)Cd_c(x),\epsilon^{abc}
u_b^T(x)Cs_c(x),\epsilon^{abc} d_b^T(x)Cs_c(x)  $ have spin-parity
 $J^P=0^+$ and  $J^P=0^-$ respectively. They
both belong to the antitriplet $\bar{3}$ representation of the color
$SU(3)$ group.  The one-gluon exchange force and the instanton
induced force can lead to significant attractions between the quarks
in the $0^+$ channels \cite{GluonInstanton}.  As the instanton
induced force results in strong attractions in the scalar diquark
channel and strong repulsions in the pseudoscalar diquark channel,
the contributions from the second and third term in Eq.(11) are
canceled.  }.

  The  calculation of  operator product expansion in the  deep Euclidean space-time region is
  straightforward and tedious, here technical details are neglected for simplicity,
  once  the analytical  results are obtained,
  then we can express the correlation function  at the level of quark-gluon
degrees of freedom into the following form through dispersion
relation,
  \begin{eqnarray}
  \Pi(p)=\gamma \cdot p \frac{1}{\pi}\int_{m_s^2}^{s_0}ds
  \frac{{\rm Im}[A(s)]}{s-p^2}+ \frac{1}{\pi}\int_{m_s^2}^{s_0}ds
  \frac{{\rm Im}[B(s)]}{s-p^2}+\cdots \, ,
  \end{eqnarray}
where
\begin{eqnarray}
\frac{{\rm Im}[A(s)]}{\pi}&=&
\frac{s^5}{2^{10}5!5!7\pi^8}+\frac{m_s \langle
\bar{s}s\rangle s^3}{2^{8}5!3!\pi^6}-\frac{m_s \langle
\bar{s}g_s \sigma  G s\rangle
s^2}{2^{9}4!3!\pi^6}+
\frac{s^3}{2^{10}5!3!\pi^6}\langle \frac{\alpha_s
GG}{\pi}\rangle \nonumber\\
&+&\frac{9\bar{n}}{{m^*}^4 } \frac{d}{dt}\left\{
\frac{s^4 t^{10}}{5!5!2^{11}\pi^6 \rho^2_c} \int_0^1 d
\alpha \int_0^1 d \beta \beta^4 (1-\beta) J_{10}
  \right.\nonumber\\
&+& \frac{m_s\langle \bar{s}s \rangle
s^3 t^8}{5!5!2^{12}3\pi^4 } \int_0^1 d \alpha
\frac{J_{8}}{\alpha(1-\alpha)}  \nonumber\\
&-& \left.\frac{m_s\langle \bar{s}g_s\sigma G s \rangle
s^{\frac{5}{2}} t^7\rho_c}{5!5!2^{15}3^2\pi^4 } \int_0^1 d \alpha \frac{ 18
\sqrt{\alpha(1-\alpha) } J_{7}
-\rho_c s^{\frac{1}{2}}t J_{8}}{(\alpha(1-\alpha))^2} \right\}|_{t=1} \, ;\nonumber\\
 \frac{{\rm Im}[B(s)]}{\pi}&=& \frac{m_s s^5}{2^{10}5!5!\pi^8}-\frac{ \langle
\bar{s}s\rangle s^4}{2^{9}5!3!\pi^6}+\frac{ \langle \bar{s}g_s
\sigma  G s\rangle s^3}{2^{9}4!3!\pi^6} \nonumber\\
&+& \frac{9\bar{n}}{{m^*}^4}\frac{d}{dt}\left\{ \frac{m_s
s^{\frac{7}{2}}t^9}{5!5!2^{11}\pi^6 \rho_c } \int_0^1 d \alpha
\int_0^1 d \beta \sqrt{\frac{\beta}{ \alpha(1-\alpha)} }\beta^4
J_{9}
  \right.\nonumber\\
&-& \frac{\langle \bar{s}s \rangle s^3 t^8}{5!5!2^{10}3\pi^4 }
\int_0^1 d \alpha
\frac{J_{8}}{\alpha(1-\alpha)}  \nonumber\\
&+& \left.\frac{m_s\langle \bar{s}g_s\sigma G s \rangle
s^{\frac{5}{2}} t^7 \rho_c}{5!5!2^{14}3\pi^4 } \int_0^1 d \alpha \frac{
18  \sqrt{\alpha(1-\alpha) } J_{7}
-\rho_c s^{\frac{1}{2}} t J_{8} }{(\alpha(1-\alpha))^2} \right\}|_{t=1} \, , \nonumber
\end{eqnarray}
here, the
$J_{10}=J_{10}\left(t\rho_c\sqrt{\frac{s\beta}{\alpha(1-\alpha)}}\right)$,
$J_{9}=J_{9}\left(t\rho_c\sqrt{\frac{s\beta}{\alpha(1-\alpha)}}\right)$,
$J_{8}=J_{8}\left(t\rho_c\sqrt{\frac{s}{\alpha(1-\alpha)}}\right)$
and
$J_{7}=J_{7}\left(t\rho_c\sqrt{\frac{s}{\alpha(1-\alpha)}}\right)$
are Bessel functions. We perform  the operator product expansion
up to the condensates of dimension 6, neglect the terms $m_s \langle
\frac{\alpha_s GG}{\pi}\rangle$ due to their small contributions.
 There are no contributions  proportional to $\langle \bar{q}q\rangle^2$ from the first term in Eq.(11),
 the direct instanton contributions from the second and third term in Eq.(11) are canceled
  due to the special interpolating current.

Matching Eq.(6) with Eq.(13) below the threshold $s_0$, then preform the Borel transform with respect to the
variable $P^2=-p^2$, we obtain the  sum rules,
\begin{eqnarray}
\lambda^2  e^{ -\frac{m^2_{\Theta^+}}{M^2} } &=& \frac{1}{\pi}\int_{m_s^2}^{s_0}ds e^{-{\frac{s}{M^2}}}{\rm Im}[A(s)], \\
\lambda^2 m_{\Theta^+}  e^{ -\frac{m^2_{\Theta^+}}{M^2} } &=& \frac{1}{\pi}\int_{m_s^2}^{s_0}ds e^{-{\frac{s}{M^2}}}{\rm Im}[B(s)].
\end{eqnarray}
Differentiate  the above sum rules with respect to the variable
$\frac{1}{M^2}$, then eliminate the quantity $\lambda^2$,
\begin{eqnarray}
m^2_{\Theta^+}&=& \frac{\int_{m_s^2}^{s_0}ds e^{-{\frac{s}{M^2}}}s{\rm Im}[A(s)]}
{\int_{m_s^2}^{s_0}ds e^{-{\frac{s}{M^2}}}{\rm Im}[A(s)]}, \\
m^2_{\Theta^+}&=& \frac{\int_{m_s^2}^{s_0}ds e^{-{\frac{s}{M^2}}}s{\rm Im}[B(s)]}
{\int_{m_s^2}^{s_0}ds e^{-{\frac{s}{M^2}}}{\rm Im}[B(s)]}.
\end{eqnarray}

 In this article, we have not shown the contributions from the higher
resonances and  continuum states explicitly for simplicity.

\section{Numerical Results}
The parameters are taken as $\langle \bar{s}s \rangle=0.8\langle
\bar{u}u \rangle$, $\langle \bar{s}g_s\sigma  G s
\rangle=m_0^2\langle \bar{s}s \rangle$, $m_0^2=0.8GeV^2$, $\langle
\bar{u}u \rangle=\langle \bar{d}d \rangle=(-0.219 GeV)^3$, $\langle
\frac{\alpha_sGG}{\pi} \rangle=(0.33GeV)^4$,
 $\bar{n}=\bar{n}_I+\bar{n}_A=1{\rm fm}^{-4}$,
$ \rho_c=\frac{1}{3}{\rm fm}$ , $m^*=170MeV$, $m_u=m_d=0$ and
$m_s=150MeV$. As the sum rules are relatively sensitive to the
condensates concerning the s quark, here we use the standard values
and neglect the  uncertainties.  The threshold parameter $s_0$ is
chosen to  vary between $\sqrt{s_0}=(1.6-2.1) GeV$ to avoid possible
pollutions from
  higher resonances and continuum states.
For the conventional ground state mesons and baryons, due to the
resonance dominates over the QCD continuum contributions, the good
convergence of the operator product expansion, and the useful
experimental guidance on the threshold parameter $s_0$, we can
obtain the fiducial Borel mass region. However, in the QCD sum rules
for the pentaquark states, the spectral density $\rho(s) \sim s^m$
with $m$ larger than  the corresponding ones in the
 sum rules for the conventional baryons, larger $m$ means   stronger dependence on the
continuum or the threshold parameter $s_0$ . Due to the large
continuum contributions,  the threshold parameter  $s_0$ has  to be
fixed ad hoc or intuitively \cite{Narison04}. In this article, the
threshold parameter $s_0$ is taken to be $\sqrt{s_0}=(1.6-2.1)GeV$,
the mass $m_{\Theta^+}=1540MeV$ and the width
$\Gamma_{\Theta}<10MeV$, the contributions from the lowest
pentaquark state can be successfully taken into account.
  In the region $M^2=(1.4-3.0)GeV^2$,
  we can obtain stable sum rules for the mass $m_{\Theta^+}$ from Eq.(16);
    no reliable sum rules can be obtained from Eq.(17). The numerical results are shown in Table 1.
 From Eq.(16), we obtain the values $m_{\Theta^+}\approx(1450-1760)MeV$ without the direct instantons;
 by including the contributions from  the direct instantons,
  we can obtain  the  mass $m_{\Theta^+}\approx (1430-1780)MeV$ for $\sqrt{s_0}=(1.7-2.0) GeV$. The contributions from
  the direct instantons are very small and can be safely neglected.
     The contributions from the direct instantons can improve the QCD sum rule greatly
 in some channels,  for example,  the
nonperturbative contributions from the direct instantons to the
conventional operator product expansion can significantly improve
the stability of chirally odd  nucleon sum rules
\cite{Dorokhov90,Forkel}.
\begin{table}[ht]
         \caption{The values of $m_{\Theta^+}$ with $M^2=2.2GeV^2$}
         \begin{center}
         \begin{tabular}{c||c||c}
         \hline\hline
         $\sqrt{s_0}$   & $m_{\Theta^+}$ $(MeV)$ & $m_{\Theta^+}$ $(MeV)$ \\
            $(GeV)$     &  With instanton        & Without instanton \\ \hline\hline
           1.6          &    1369                 &  1395    \\ \hline
           1.7          &    1464                 &  1483    \\ \hline
           1.8          &    1567                 &  1571    \\ \hline
           1.9          &    1670                 &  1660    \\ \hline
           2.0          &    1769                 &  1748   \\ \hline
           2.1          &    1861                 &  1836   \\  \hline\hline
         \end{tabular}
         \end{center}
         \end{table}

\section{Conclusion }

In this article, we take the point of view that the pentaquark state $\Theta^+(1540)$
 have quantum numbers, $J=1/2$, $I=0$, $S=+1$ and study
 its mass with the (scalar)diquark-(pseudoscalar)diquark-antiquark type interpolating
  current in the framework of  the QCD sum rules approach
by including the contributions from the direct instantons. As the
instanton vacuum is fairly dilute,  we can take the single instanton
approximation mathematically, the collective contributions  from
all
 instantons other than the leading one
are taken into account  by a single effective parameter, the
effective mass  $m^*$. The numerical results indicate that the
contributions from  the direct instantons are very small and can be
safely neglected.

\section*{Acknowledgment}
This  work is supported by National Natural Science Foundation,
Grant Number 10405009,  and Key Program Foundation of NCEPU. The
authors are indebted to Dr. J.He (IHEP), Dr. X.B.Huang (PKU) and Dr. L.Li (GSCAS)
for numerous help, without them, the work would not be finished.

\end{document}